\documentclass[11pt,a4paper]{article}
\pdfoutput=1 

\usepackage{jheppub}
\usepackage{color}
\usepackage{textcomp}
\usepackage{multirow}
\usepackage{bm}
\title{Excited spectroscopy of charmed mesons from lattice QCD}

\author[a]{Graham~Moir,} \emailAdd{moirg@tcd.ie}
\author[a]{Michael~Peardon,} \emailAdd{mjp@maths.tcd.ie}
\author[a]{Sin\'ead~M.~Ryan,} \emailAdd{ryan@maths.tcd.ie}
\author[a]{Christopher~E.~Thomas} \emailAdd{thomasc@maths.tcd.ie}  

\author[b]{and Liuming~Liu} \emailAdd{liuming@hiskp.uni-bonn.de}

\affiliation[a]{School of Mathematics, Trinity College, Dublin~2, Ireland}
\affiliation[b]{Helmholz-Institut f\"ur Strahlen-und Kernphysik and Bethe Center for Theoretical Physics, Universit\"at Bonn, D-53115 Bonn, Germany}

\affiliation{\vspace{0.1cm} {\rmfamily \normalsize (for the Hadron Spectrum Collaboration)}}

\preprint{TCDMATH 13-01}

\abstract{We present spectra of highly excited $D$ and $D_{s}$ mesons up to around $3.8$~GeV determined 
using dynamical lattice QCD. We employ novel computational 
techniques and the variational method with a large basis of carefully constructed operators in order to 
extract and reliably identify the continuum spin of an extensive set of excited states. These include states with high spin and states identified as having an explicit gluonic contribution. 
Calculations were performed on two volumes, both with a pion mass of
approximately $400$ MeV, achieving a high 
statistical precision for both ground and excited states. We discuss our results in light of
 experimental observations, comment on the phenomenological implications and identify the lightest `supermultiplet' of hybrid mesons in each sector.}

\begin{document}
\maketitle

\section{Introduction}
\label{sec:introduction}

The spectra of open-charm mesons contain a number of experimentally
well-established states~\cite{PDG:2012}: eight ``charm-light'' $D$ mesons with isospin $I=1/2$ and six ``charm-strange'' $D_s$
mesons with non-zero strangeness and zero isospin. 
Many of the states whose spin, $J$, and parity, $P$, are well determined fit the pattern expected from model predictions 
but a subset of these, as well as additional states which require experimental confirmation, do not and remain unexplained.  
Suggestions such as hybrid mesons (in which the gluonic field is excited), molecular mesons and tetraquarks 
(with four valence quarks) have been proposed, but no clear picture has emerged.
In particular, the masses and widths of the enigmatic 
$D^{*}_{s0}(2317)^{\pm}$ and $D_{s1}(2460)^{\pm}$ are lighter and narrower than in quark models~\cite{Godfrey:1985,Close:2005}. 
Since the discovery of these resonances at BABAR~\cite{BABAR:2003} and CLEO~\cite{CLEO:2003} respectively, 
there has been much debate about their nature and the situation is yet to be resolved.

A simple non-relativistic picture of a meson describes it as a bound state of a quark and an 
antiquark whose spins are coupled together to form a total spin $S$, which is then coupled to an orbital angular momentum $L$
in order to create a state of total angular momentum $J$. 
If the meson is an eigenstate of charge-conjugation with quantum number $C$, 
some $J^{PC}$ combinations are inaccessible in this description; 
such states are termed `exotic'.  The charmonium system provides an arena
for experiments such as BESIII and PANDA to search for these exotics, with
the observation of such a state being a smoking gun for physics beyond quark models.
A recent study of charmonium by the Hadron Spectrum Collaboration found states with exotic quantum numbers and showed that these, along with other non-exotic states, could be identified as hybrid mesons~\cite{Liu:2012}.
On the other hand, open-charm mesons are not eigenstates of charge conjugation and so there are no such exotic quantum number combinations. This increases the difficulty of experimentally identifying a multiquark or hybrid state in the open-charm sector. However, hybrid states are still expected to be present in the spectrum and the system provides an important probe of the dynamics of excited 
glue in another environment. 

A full understanding of QCD and its spectrum requires ab initio calculations to 
compare against and inform experimental searches, and to discriminate between 
models to glean the relevant physics. 
Spectroscopic calculations on the lattice have a long history 
and an accurate determination of the low-lying experimental spectrum is considered an acid test of the precision and accuracy of the method. 
Current lattice QCD calculations have reached this benchmark as shown by the extremely precise calculations of the 
so-called `gold plated' hadron spectrum~\cite{Durr:2008}.

The extraction of higher-lying spectroscopic states on the lattice is more problematic. 
In lattice calculations space-time is discretised on a hypercubic grid of points and its volume reduced to that of a few fermi. 
The breaking of Lorentz invariance reduces the full continuum rotation symmetry to that of the octahedral group, $O_{h}$; this means that states at rest are no longer classified according to their spin, $J$, but by the irreducible representations (\emph{irreps}) of the octahederal group. Since there are only five irreps\footnote{here we are only considering the single-cover representations relevant for mesons} of $O_{h}$ for each parity and an infinite number of spins, each irrep contains an infinite number of spins and the various components of states with $J \geq 2$ are distributed across several irreps; this makes the task of disentangling the spin of states difficult. The Hadron Spectrum Collaboration has addressed the problem of 
this reduced symmetry through the use of large bases of carefully constructed interpolating operators. In combination with dynamical anisotropic lattices~\cite{Edwards:2008,Lin:2009} and the distillation technique~\cite{Peardon:2009}, this methodology has recently been used successfully to extract highly excited spectra and to reliably identify the continuum quantum numbers of high spin states for both light and heavy mesons~\cite{Dudek:2009,Dudek:2010,Dudek:2011,Liu:2012}, as well as for baryons~\cite{Bulava:2010,Edwards:2011,Dudek:2012,Edwards:2012fx}. We will use the same methodology in this study and summarise the important points below.

Previous calculations of the open-charm meson spectra have mainly focused on determining the lowest-lying $S$-wave [$(0,1)^-$] and $P$-wave [$(0,1,2)^+$] states (see Refs.~\cite{Namekawa:2011wt,McNeile:2012qf,Dowdall:2012} for some recent investigations).  Only more recently have calculations begun to explore higher up in the spectrum~\cite{Mohler:2011,Mohler:2012na,Bali:2011,Bali:2012} and we discuss these studies in Section~\ref{subsec:final spectra}. In this paper we extract $D$ and $D_{s}$ meson excitation spectra at a single lattice spacing with a pion mass of approximately $400$ MeV and we reliably identify the continuum quantum numbers of the extracted states. The results presented here are the most extensive determination of the open-charm spectra from lattice QCD to date, include all $J^P$ combinations with $J \leq 4$ and consider states with a hybrid nature. Preliminary results from this work have been presented in conference proceedings~\cite{Moir:2013ej}.

In addition, this work lays a foundation for more detailed investigations of
resonances above strong-decay threshold in both charmonium and the 
open-charm sector, including the enigmatic new states.
These studies will consider scattering involving $D$ and $D_s$ mesons for which the 
determination of masses and dispersion relations discussed in this paper
is an important prerequisite. We can compute the multi-meson energy levels 
in a finite volume by supplementing our operators with those designed to efficiently create
such states~\cite{Dudek:2012a,Dudek:2012xn}. 
The resulting denser spectrum can be analysed, at least in the case of elastic scattering,
using the L\"uscher methodology~\cite{Luscher:1990} and its extensions, and the mass and width of resonances determined.

\section{Lattice discretisation and interpolating operators}
\label{sec:lattice calculation}

The calculations presented here make use of ensembles generated by the Hadron 
Spectrum Collaboration; for more details see 
Refs.~\cite{Edwards:2008,Lin:2009}. These gauge field configurations include
the dynamics of two mass-degenerate light quarks and a strange quark, while
the charm-quark field is quenched. 
We use an anisotropic discretisation in which the spatial lattice spacing, $a_{s}$, and the temporal lattice spacing, $a_{t}$, are distinct and related via 
$\xi = a_{s} / a_{t} \approx 3.5$. This ensures that we have
$a_{t}m_{c} \ll 1$, where $m_c$ is the bare charm-quark mass, 
and the standard relativistic formulation of fermions can be used to 
study states in which the charm quark four-momentum is closely aligned with the
temporal axis. 
It is worth noting that in these computations $a_{s}m_{c}$ is also less than
unity.  The anisotropy, $\xi$, is discussed further in Section~\ref{sec:dispersion_relations}.

\subsection{Gauge and light-quark fields}
\label{subsec:actions}

In the gauge sector, a Symanzik-improved anisotropic action with tree-level
tadpole-improved coefficients is used. 
For all the quark fields, a tree-level tadpole-improved Sheikholeslami-Wohlert
(clover) anisotropic action with stout-smeared spatial links~\cite{Sheikholeslami:1985,Morningstar:2003} is employed. 
In Refs.~\cite{Edwards:2008,Lin:2009} the parameters of these actions were
determined in order to study
light-hadron spectroscopy using dynamical light and strange quarks. 

In lattice computations all quantities are calculated in terms of the lattice spacing
and, to make contact with experiment and present results in physical units,
a scale-setting is introduced. 
Here we follow Ref.~\cite{Lin:2009}, using the ratio of the $\Omega$-baryon mass 
measured on these ensembles, $a_{t}m_\Omega = 0.2951(22)$~\cite{Edwards:2011}, to the experimental mass, $m_\Omega = 1672.45(29)$ MeV~\cite{PDG:2012},
to determine $a_t$. 
This yields $a^{-1}_{t} = 5.67(4)$~GeV and a spatial lattice spacing about $3.5$ times larger, $a_s \approx 0.12$~fm.
We perform calculations on two lattice volumes
corresponding to spatial extents of about $1.9$~fm and $2.9$~fm. Table~\ref{tab:latt_details} summarises the two ensembles 
used in this study, with full details given in Refs.~\cite{Edwards:2008,Lin:2009}.

\begin{table}[tb]
\begin{center}
\begin{tabular}{ccccc}
Lattice size & $m_\pi/$MeV & $N_{\rm cfgs}$ & $N_{\rm tsrcs}$ & $N_{\rm vecs}$ 
   \\
\hline
$16^3\times 128$ & 391 &  96 & 128 & 64  \\
$24^3\times 128$ & 391 & 553 &  16 & 162
\end{tabular}
\caption{The gauge-field ensembles and quark propagators used in this work 
on two lattice volumes, $(L/a_s)^3 \times (T/a_t)$, where $L$ and $T$ are 
respectively the spatial and temporal extents of the lattice.
The number of gauge-field configurations used, $N_{\rm cfgs}$, and the number
of time-sources for quark propagators per configuration, $N_{\rm tsrcs}$, are shown;
$N_{\rm vecs}$ refers to the number of eigenvectors used in the distillation method~\cite{Peardon:2009}. }
\label{tab:latt_details}
\end{center}
\end{table}

\subsection{The charm-quark action}

Lattice QCD has a long-standing problem when trying to perform calculations
involving heavy quarks moving relativistically. 
One requires very fine lattices because
the usual lattice actions lead to large discretisation artefacts when the 
lattice spacing $a \geq m^{-1}_{0}$, where $m_{0}$ is the bare
quark mass. In theory there is a simple solution: perform calculations on 
a much finer lattice. But in practice, this is computationally expensive 
if a large enough spatial volume is maintained.
Recently, anisotropic lattices have been used 
to circumvent the problem by employing a temporal lattice spacing smaller than 
that in the spatial direction. This enables simulations with $a_{t}m_{0} \ll 1$, 
allowing correlation functions to be studied in detail and 
discretisation artefacts of heavy quarks
to be kept under control, while maintaining the spatial directions 
coarse enough to keep the computational cost reasonable.

In Ref.~\cite{Liu:2012}, a valence charm quark was introduced and used to 
study the charmonium spectrum.  The bare mass of the charm quark was non-perturbatively 
determined by requiring the ratio of the $\eta_{c}$ mass to the 
$\Omega$-baryon mass to take its experimental value. 
The same action was used for the charm-quark field as for
the light and strange quarks, except the parameter giving relative weights to 
spatial and temporal finite differences was tuned to
give a relativistic dispersion relation, consistent with $\xi = a_s/a_t = 3.5$,
for the $\eta_c$ meson at low momentum.
In this study we use the same actions and set of parameters.  
In Section~\ref{sec:dispersion_relations}, we investigate $D$ and $D_s$ mesons 
with non-zero momentum and find that they satisfy a relativistic dispersion 
relation, again consistent with $\xi = 3.5$.

\subsection{Operator construction}
\label{subsec:operator construction}

Spectral information can be obtained from Euclidean two-point correlation 
functions determined by lattice QCD computations,
\begin{equation}
C_{ij}(t)= \langle 0|{\cal O}_i(t){\cal O}^{\dagger}_j(0)|0\rangle, 
\end{equation}
where ${\cal O}^{\dagger}(0)$ and ${\cal O}(t)$ are respectively the source and sink interpolating fields (creation and annihilation operators).
By inserting a complete set of eigenstates of the Hamiltonian, the correlation 
function becomes a sum over all states with the same quantum numbers of the 
interpolating fields,
\begin{equation}
C_{ij}(t) = \sum_\mathfrak{n} \frac{Z_i^{\mathfrak{n}*} Z_j^{\mathfrak{n}}}{2E_\mathfrak{n}} e^{-E_\mathfrak{n}t} ~,
\end{equation}
with the discrete nature of the spectrum arising from the finite volume of the lattice. The vacuum-state matrix elements
$Z^{\mathfrak{n}}_i \equiv \langle \mathfrak{n} | \mathcal{O}^{\dagger}_i | 0 \rangle$, or \emph{overlaps}, are used to identify
the continuum spin of the extracted states as discussed in Section~\ref{subsec:spin identification} and to probe their structure.

\begin{table}[tb]
\begin{center}
\begin{tabular}{l l | cccccccc}
         & &$a_0$ &$\pi$ &$\pi_2$ &$b_0$ &$\rho$ &$\rho_2$ &$a_1$ &$b_1$ \\
\hline
$\Gamma$ & &1 &$\gamma_5$ &$\gamma_0 \gamma_5$ &$\gamma_0$ &$\gamma_i$ &$\gamma_0 \gamma_i$ &$\gamma_5 \gamma_i$ & $\gamma_0 \gamma_5 \gamma_i$ 
\end{tabular}
\caption{Gamma matrix combinations and naming scheme.}
\label{table:GammaMatrix}
\end{center}
\end{table}

In order to maximise the spectral information we can extract from correlation functions we employ the following
strategies. Firstly, we construct a large basis of operators with a variety of spin and spatial structures
 that allows us to explore all $J^{P}$ combinations with $J \leq 4$. 
In general, these operators take the form of a quark bilinear 
with an operator insertion $\mathbf \Gamma$,
\begin{equation}
{\cal O}_{f\!f'} = \bar{\psi}_f\; {\mathbf \Gamma}\; \psi_{f'} = 
\bar{\psi}_f\;\Gamma \overleftrightarrow{D}_i \overleftrightarrow{D}_j
\cdots\psi_{f'} ~,
\end{equation}
where $f$,$f'$ label the quark flavours, $\Gamma$ is a gamma matrix 
combination taken from Table~\ref{table:GammaMatrix} and the projection onto 
definite momentum and
spatial indices are suppressed for clarity. 
The `forward-backward' covariant derivative is defined as 
$\overleftrightarrow{D} \equiv \overleftarrow{D} - \overrightarrow{D}$.
As described in Refs.~\cite{Dudek:2009,Dudek:2010}, the derivatives and vector-like gamma matrices are first expressed in a circular basis so that they transform as $J=1$. Using the usual $SO(3)$ Clebsch-Gordan coefficients they can then be combined to form operators, $\mathcal{O}^J$, with any desired continuum $J^P$. In order to use these operators in lattice calculations they are \emph{subduced} into the relevant irrep(s) of $O_h$~\cite{Dudek:2010} and Table~\ref{Table:Subduce} shows how the continuum spins are distributed across these irreps. Operators subduced into irrep $\Lambda$ and irrep row $\lambda$ from spin $J$ are denoted by $\mathcal{O}_{\Lambda,\lambda}^{[J]}$ and our operator naming scheme is explained in Ref.~\cite{Dudek:2010}. The number of subduced operators employed in each irrep is shown in Table~\ref{table:ops_per_irrep}.

\begin{table}[tb]
\begin{center}
\begin{tabular}{ccl}
$J$ & & $\Lambda(\text{dim})$ \\
\hline
$0$ & & $A_1(1)$ \\
$1$ & & $T_1(3)$ \\
$2$ & & $T_2(3) \oplus E(2)$\\
$3$ & & $T_1(3) \oplus T_2(3) \oplus A_2(1)$\\
$4$ & & $A_1(1) \oplus T_1(3) \oplus T_2(3) \oplus E(2)$
\end{tabular}  
\caption{Distribution of continuum spins across the irreps of $O_{h}$, $\Lambda(\text{dim})$, where dim is the dimension of the irrep.}
\label{Table:Subduce}
\end{center}
\end{table}

\begin{table}[tb]
\begin{center}
\begin{tabular}{l l|cc}
$\Lambda$ & & $\Lambda^{-}$ & $\Lambda^{+}$ \\
\hline
$A_1$     & & 18   &  18 \\
$A_2$     & & 10   &  10 \\
$T_1$     & & 44   &  44 \\
$T_2$     & & 36   &  36 \\
$E$       & & 26   &  26 \\
\end{tabular}
\caption{The number of operators used in each lattice irrep, $\Lambda^{P}$. All combinations of gamma matrices and up to three derivatives are included.}
\label{table:ops_per_irrep}
\end{center}
\end{table}

The operator in the bilinear, $\mathbf \Gamma$, is constructed to have 
well-defined transformation properties under transposition,
${\mathbf \Gamma}^T = \pm {\mathbf \Gamma}$. We
denote the eigenvalue of $\mathbf \Gamma$ under this transformation as 
$C_\Gamma=\pm 1$. If the quark and antiquark fields 
have the same flavour, {\it i.e.} $f=f'$, this corresponds to
the charge-conjugation quantum number, $C$. 
Similarly, if $f \neq f'$ but $f$ and $f'$ are degenerate in mass,
this is related to a quantum number which is a generalisation of $C$, e.g. $G$-parity 
relevant for mesons consisting of degenerate up and down quarks.
For $D$ and $D_s$ mesons, $f=c$, $f'\in \{u,d,s\}$ and 
there is no mass degeneracy between the quark and antiquark fields in our operators.
As a result, states created by operators with opposite $C_\Gamma$
can mix to form an eigenstate of the QCD Hamiltonian.  This effect will be
considered in more detail in Section~\ref{sec:analysis}.
   
Secondly, we employ a now well-established quark smearing procedure known as \emph{distillation}~\cite{Peardon:2009}. This procedure consists of applying
a smoothing function to the quark fields, reducing the contamination from noisy UV modes
that do not make a significant contribution to the low-energy physics we wish to
extract. The distillation operator is defined via,
\begin{equation}
 \Box_{ij}(t) = \sum^{N_{vecs}}_{n=1} f(\lambda^{(n)}) v^{(n)}_{i}(t)
      v^{(n)\dagger}_{j}(t),
\end{equation}
where $v^{(n)}_{i}(t)$ are the eigenvectors, sorted by eigenvalue $\lambda^{(n)}$, of the three-dimensional gauge-covariant lattice Laplacian evaluated on the background of spatial 
gauge fields of timeslice $t$. $f(\lambda^{(n)})$ is a smearing profile and in this study it is
set to unity, although other smearing profiles could prove to be beneficial. Apart from the removal of short-distance modes, distillation has other significant benefits. One advantage arises from the outer product structure of the distillation operator, $\Box$; the correlation function can be factorised into 
\emph{perambulators} and \emph{elementals} with the former containing 
information on the propagation of the quarks and the 
latter encapsulating the momentum and structure of the state created.
This factorisation allows the perambulators to be stored and so
reduces the computational cost of the calculation. Once the linear systems
arising from the Dirac equation on a given gauge background have been
solved for as many time-sources as required, correlation functions involving 
any elementals can be computed without further inversions.
This enables us to perform calculations with large bases of operators 
having a variety of spin and spatial structures.
In addition, distillation enables the efficient computation of annihilation
 contributions and of correlation functions involving multi-hadron operators.

Using distillation, for each irrep we compute two-point correlation functions 
for all possible combinations of our operators. Our methodology for analysing 
these correlators is described in the following section.

\section{Correlator analysis}
\label{sec:analysis}

For each lattice irrep and flavour sector ($D$ and $D_s$) we compute an $N \times N$ matrix of correlation functions, 
\begin{equation}
 C_{ij}(t)= \langle 0|{\cal O}_i(t){\cal O}^{\dagger}_j(0)|0\rangle ,
\end{equation}
where $i,j$ label the operators in the basis and $N$ is the number of operators in the irrep, given in Table~\ref{table:ops_per_irrep}.  We then apply the variational method~\cite{Luscher:1990,Michael:1985}, now commonly used in lattice QCD calculations of spectra, which finds the optimal extraction, in a variational sense, of energies in a given channel.  The details of our particular implementation along with various systematic tests are given in Ref.~\cite{Dudek:2010} and tests in the charmonium sector have been presented in Ref.~\cite{Liu:2012}; here we will briefly summarise the important points.  Practically, the procedure is realised via the solution of a generalised eigenvalue problem,
\begin{equation}
 C_{ij}(t) v^{\mathfrak{n}}_j = \lambda^{\mathfrak{n}}(t,t_0) C_{ij}(t_0)
    v^{\mathfrak{n}}_j, 
\end{equation}
where the reference time-slice $t_{0}$ must be chosen appropriately as described in Refs.~\cite{Dudek:2007,Dudek:2010}.   Solving this for each $t$ yields eigenvalues, $\lambda^{\mathfrak{n}}(t,t_0)$, and eigenvectors, $v^{\mathfrak{n}}_j$, where $\mathfrak{n} = 1,2,\ldots,N$.  The method is very powerful for extracting excited-state energies; even nearly degenerate states can be distinguished because such states in general have different patterns of operator-state overlaps corresponding to distinct orthogonal eigenvectors. 

The eigenvalues, known as \emph{principal correlators}, allow access to the
energies, $E_{\mathfrak{n}}$, via their dependence on $(t - t_{0})$; for
sufficiently large times they are proportional to $e^{-E_{\mathfrak{n}}(t-t_{0})}$.  The eigenvectors are related to the overlaps, $Z^{\mathfrak{n}}_i$, which give information on the structure of a state and, as will be described below, enable us to identify the continuum spin of a state.  For insufficiently large times, there will be some contamination from excited states in the principal correlators.  In order to `mop up' this contamination and enable us to stabilise the fits by considering a larger temporal extent, we use a fitting function containing two exponentials of the form,
\begin{equation}
 \lambda_\mathfrak{n}(t) = (1 - A_\mathfrak{n}) e^{-m_\mathfrak{n}(t-t_0)} + A_\mathfrak{n} e^{-m_\mathfrak{n}' (t-t_0)}~,
\end{equation}
where the fit parameters are $m_\mathfrak{n}$, $m_\mathfrak{n}'$ and $A_\mathfrak{n}$.  Empirically we find that the contribution from the second exponential decreases rapidly as we increase $t_0$.  Our best estimate of the mass of the state is $m_\mathfrak{n}$ and we do not further consider the quantities $m_\mathfrak{n}'$ and $A_\mathfrak{n}$.

The combination of distillation and variational analysis of correlator matrices with large bases of subduced operators has proved very effective at extracting spectra of excited and high-spin light mesons~\cite{Dudek:2009,Dudek:2010,Dudek:2011}, charmonia~\cite{Liu:2012} and baryons~\cite{Edwards:2011,Dudek:2012}.

\subsection{Spin identification}
\label{subsec:spin identification}

In principle, to determine the continuum spin, $J$, of states extracted in
lattice QCD calculations, the spectrum can be computed at finer and finer lattice
spacings and then extrapolated to the continuum limit.  Exact degeneracies will
then emerge between states of the same spin subduced into different irreps of
$O_{h}$; these can be matched up between irreps enabling the spin to be
determined from the pattern of degeneracies.  However, there are a number of
problems with this approach.  The high computational cost of performing
calculations at multiple fine lattice spacings means that this is not currently 
feasible for computations of excited states.
More problematically, there can be physical degeneracies
or near-degeneracies in the dense high-lying spectrum; the question then arises 
as to how to distinguish these, with a finite statistical precision, 
from artefacts due to a finite lattice spacing.
It would be useful to have a spin identification scheme that can be used to
reliably determine the spin at a single lattice spacing.  As we will see in
Section~\ref{sec:results}, the dense spectrum of excited states would be
impossible to disentangle without more information beyond solely the masses of
states.  This is particularly relevant in the $D$ and $D_s$ meson sectors where
a lack of charge-conjugation symmetry makes the spectrum more dense than, for 
example, the charmonium sector.  

The Hadron Spectrum Collaboration has recently developed a
scheme~\cite{Dudek:2009,Dudek:2010} to identify the continuum spin of an
extracted state at a single lattice spacing by using information from the
operator-state overlaps defined in Section~\ref{subsec:operator construction}.
Further details and a demonstration of the efficacy of this approach are
presented in Ref.~\cite{Dudek:2010} and here we will summarise the method; an
application to the charmonium sector is given in Ref.~\cite{Liu:2012}.  The
subduced operators, $\mathcal{O}_{\Lambda,\lambda}^{[J]}$, respect the symmetry
of the lattice but we find that they also carry a `memory' of the continuum
spin, $J$, from which they were subduced.  
If full rotational symmetry reasonably describes transformations of the 
smeared fields on hadronic scales, we would expect that an operator subduced from a continuum spin $J$ will overlap predominantly onto states of continuum spin $J$~\cite{Davoudi:2012ya}.

\begin{figure}[tb]
\begin{center}
\includegraphics[width=0.6\textwidth]{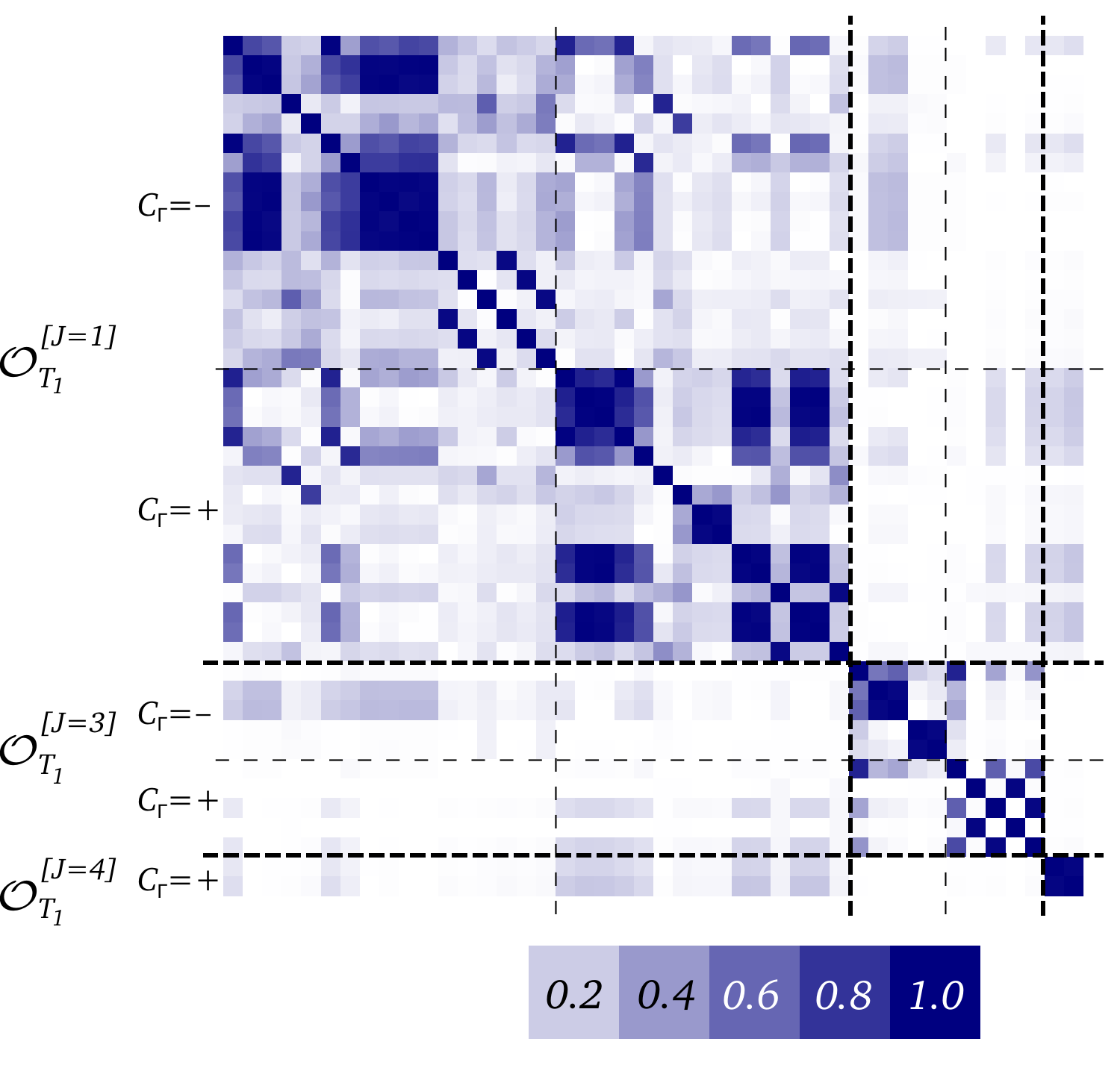}
\caption{The normalised correlation matrix, $C_{ij} / \sqrt{C_{ii}C_{jj}}$, on
  time-slice five for the $T^{+}_{1}$ irrep in the charm-light ($D$) sector on
    the $24^3$ volume.  The operators are ordered according to continuum spin,
	$J$, as indicated by the labelling.  For each value of $J$ the operators
are then ordered according to their symmetry, $C_\Gamma=\pm$, as described in the text.  The correlation matrix is observed to be approximately block diagonal in spin.} 
\label{fig:Corr_matrix} 
\end{center}
\end{figure}

\begin{center}
\begin{figure}[tb]
\includegraphics[width=1.0\textwidth]{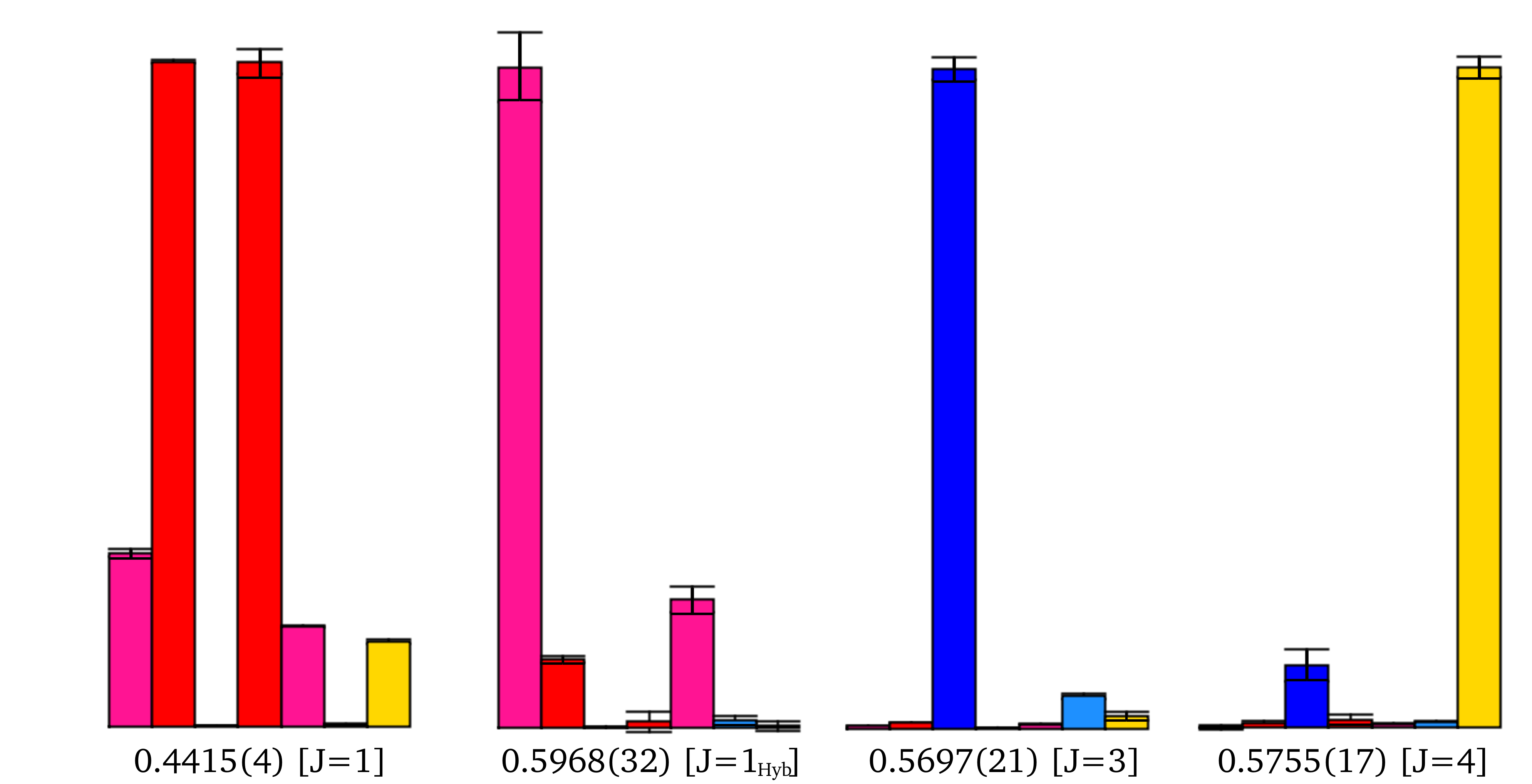}
\caption{The overlaps, $Z$, of a selection of operators onto a selection of
  states in the charm-strange ($D_s$) $T^{+}_{1}$ irrep on the $24^{3}$ volume. 
In this plot the overlaps have been normalised so that the largest value for an
operator across all states is equal to unity.  The error bars indicate the one
sigma statistical uncertainty.  For each state, the operators are coloured from
left to right as: pink $(a_{0} \times D^{[2]}_{J=1})^{[J=1]}$, red $(b_{1}
    \times D^{[2]}_{J=0})^{[J=1]}$, blue $(b_{1} \times D^{[2]}_{J=2})^{[J=3]}$,
     red $(\rho \times D^{[1]}_{J=1})^{[J=1]}$, pink $(\rho \times
	 D^{[3]}_{J_{13}=2,J=2})^{[J=1]}$, light blue $(\rho \times
	   D^{[3]}_{J_{13}=2,J=2})^{[J=3]}$, and gold $(\rho \times
	     D^{[3]}_{J_{13}=2,J=3})^{[J=4]}$. The first three operator insertions  have negative symmetry ($C_\Gamma=-$) and the last four have positive symmetry ($C_\Gamma=+$) as explained in the text. States are labelled by their mass in temporal lattice units and the continuum spin of the dominant operators. `Hyb' refers to a state which has relatively strong overlap with operators that are proportional to the field strength tensor, the commutator of two gauge-covariant derivatives.} 
\label{fig:Skyscraper}
\end{figure}
\end{center} 
\vspace{-0.95cm}

Figs.~\ref{fig:Corr_matrix} and \ref{fig:Skyscraper} show that these statements are justified at the level of the correlation matrix and the overlaps respectively.  In Fig.~\ref{fig:Corr_matrix} we show the normalised correlation matrix at time-slice 5 for the $T^{+}_{1}$ irrep in the charm-light ($D$) sector.  The matrix is approximately block diagonal in $J$, suggesting that, for example, an operator subduced from $J=1$ has little contribution from the $J=3$ and $J=4$ states.  As discussed in Section~\ref{subsec:operator construction}, charm-light and charm-strange mesons are \emph{not} eigenstates of charge conjugation.  However, our operator insertions have a definite symmetry under transposition (equivalent to charge conjugation of the operator followed by the interchange of the quark and antiquark flavours) and we use this to label the operators, $C_\Gamma=\pm$.  This would correspond to a generalisation of charge-conjugation parity (analogous to $G$-parity) if the charm and light/strange quarks were degenerate, i.e. if we had $SU(4)$ symmetry.  The lack of such a symmetry is shown in Fig.~\ref{fig:Corr_matrix} where significant overlap between $C_\Gamma=+$ and $C_\Gamma=-$ operators is observed.

In Fig.~\ref{fig:Skyscraper} we show the overlaps of a selection of operators
onto a selection of low-lying states in the $T^{+}_{1}$ irrep of the
charm-strange ($D_s$) sector.  Every state clearly shows a strong preference to
overlap onto only operators subduced from a particular spin.  The lack of a
generalised charge-conjugation symmetry is shown by the observation that the
$J=1$ state has strong overlap onto both $C_\Gamma=+$ and $C_\Gamma=-$
operators.  The subset of overlaps we show here is representative of the full
operator basis -- the observation that each state only overlaps strongly onto
operators subduced from a particular spin is repeated across our entire set of 
spectra.

\begin{center}
\begin{figure}[tb]
\includegraphics[width=1.0\textwidth]{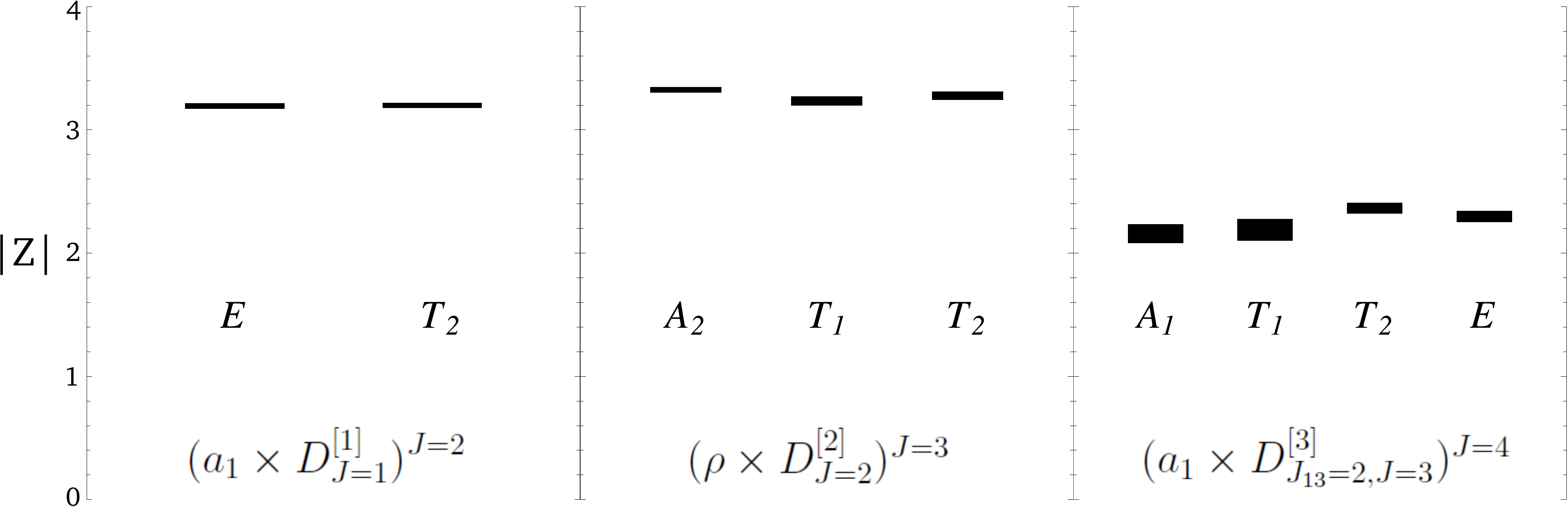}
\caption{A selection of absolute $Z$-values for the lightest states conjectured to have
  $J=2$ (left pane), $J=3$ (middle pane), $J=4$ (right pane) in charm-strange
    ($D_s$) irreps, $\Lambda^{-}$, on the $24^{3}$ volume. For each $J$ we show
    the values of $Z$ for the same operator subduced into different irreps. The vertical size of a box corresponds to the one-sigma statistical uncertainty on either side of the mean.}
\label{fig:Z_values}
\end{figure}
\end{center}

For $J \ge 2$, we can also match up the states across different irreps of
$O_{h}$ by comparing the absolute values of the overlaps. Up to discretisation
effects, these values should be consistent for states subduced from the same
continuum state with spin $J$. In Fig.~\ref{fig:Z_values} we plot a selection of
overlaps for states conjectured to have $J^{P} = 2^{-}, 3^{-}$ and $4^{-}$, demonstrating there is good agreement between the overlaps extracted in different irreps.  This observation is repeated across our entire spectra though, as we expect, there are slight deviations from equality due to discretisation effects.  To quote masses for $J \ge 2$ states we perform joint fits to the principal correlators in each irrep as described in Ref.~\cite{Dudek:2010}.

\vspace{-1.3cm}

In summary, we have demonstrated that the method for identifying the continuum spin of extracted states that was developed in Refs.~\cite{Dudek:2009,Dudek:2010} is applicable to studies of charm-light and charm-strange mesons.  We will use this approach to identify the $J$ of the extracted states that we present in Section~\ref{sec:results}.


\section{Dispersion relations}
\label{sec:dispersion_relations}

One purpose of this investigation is to ensure we can study heavy-light mesons
reliably on these anisotropic lattices to enable subsequent calculations 
involving scattering of these states. This is essential if the nature
of the charmonium and charmed resonances above the open-charm threshold is to be 
understood. To determine scattering properties of mesons using lattice QCD 
requires the computation of the full spectrum of energy eigenstates, including
those made up predominantly of more than one meson. For these determinations to
be reliable, the dispersion relations for the mesons produced by decays of 
the resonances should be relativistic up to the appropriate momenta set by
the decay kinematics. Heavy quarks bring extra complications to 
these calculations but the anisotropic discretisation we use here may reduce the
impact of the large mass scale, $m_Q$. For mesons moving with modest momenta,
the heavy quark four-momentum will predominantly be aligned with the temporal 
direction, the most finely discretised lattice direction. 

One well-studied artefact of heavy quarks on the lattice is that when
$a m_Q$ is ${\cal O}(1)$ the dispersion relation for mesons appears
non-relativistic and the kinetic and rest masses of the mesons differ.
Heavy-light and heavy-heavy states are effected differently. 
On the anisotropic lattice, the relativistic dispersion relation is most 
naturally written,
\begin{equation}
\label{equ:dispersion}
 (a_{t}E)^{2} = (a_{t}m)^{2} + \left(\frac{1}{\xi}\right)^{2} (a_{s}|\vec{p}|)^{2} ~,
\end{equation}
where $\vec{p} = \frac{2\pi}{L}\vec{n}$ follows from the periodic boundary conditions of the spatial directions. Here 
$\vec{n} = (n_{x},n_{y},n_{z})$ with $n_{i}$ taking integer values.
To determine
whether the anisotropic lattice has controlled the size of these artefacts, 
we determined the dispersion relations for a number of mesons in the $D$ and 
$D_s$ spectra.  Recall that the parameters in the action for the charm quark
are determined such that the $\eta_c$ meson has a relativistic dispersion
relation consistent with $\xi = 3.5$; we obtained $\xi_{\eta_c} = 3.50(2)$~\cite{Liu:2012}.
This compares with an anisotropy measured from the pion dispersion relation
of $\xi_\pi = 3.444(6)$~\cite{Dudek:2012a}.
If there was a problem with large discretisation artefacts, one 
symptom could be that the $\xi$ measured from the $D$ or $D_s$ dispersion 
relations would be different from the $\xi$ determined from the pion or the $\eta_c$.

\begin{figure}[t]
\centering
\includegraphics[width=\textwidth]{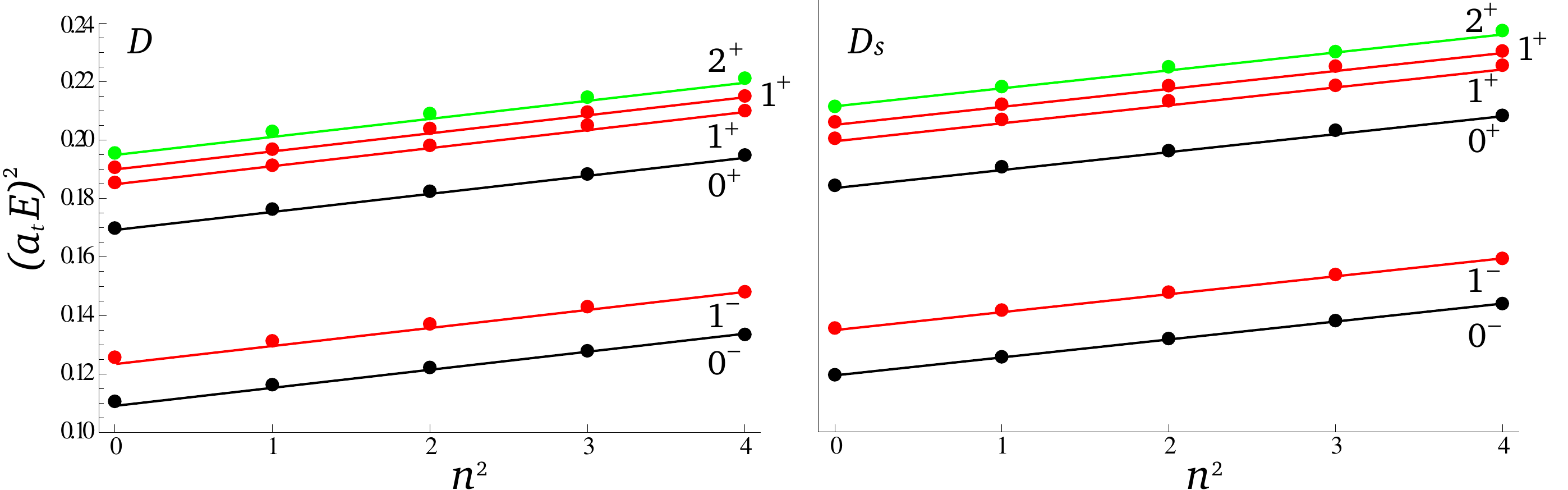} 
\caption{Squared energies as a function of $\lvert\vec{n}\rvert^{2}$ for the
lightest $S$-wave and $P$-wave states in both the $D$ (left panel) and $D_{s}$ (right panel) sectors.
The points correspond to calculated energies and the statistical uncertainties are smaller than the size of the points.
The lines are dispersion relations of the form of Eq.~\ref{equ:dispersion},
as described in the text, with $m$ fixed to the rest energy of the appropriate meson and $\xi$
fixed to $3.454$ for the charm-light mesons and $3.453$ for the charm-strange mesons.
The measured energies are seen to be well described by the dispersion relations.}
\label{fig:dispersion relations}
\end{figure}

To calculate dispersion relations for a subset of $D$ and $D_s$ mesons, we use 
two sets of operators. For mesons at rest we use the same basis of operators as described in 
Section~\ref{subsec:operator construction}, but for mesons of non-zero momenta we follow Ref.~\cite{Thomas:2012} and use operators constructed in a helicity basis (with up to two derivatives) and then subduced into the relevant irreps.
In Fig.~\ref{fig:dispersion relations} we show the squared energies as a function of $|\vec{n}|^2$
for the low-lying states in both the charm-light and charm-strange sectors. The points correspond to
calculated energies and, as usual, if the state subduces into more than one irrep, 
these energies are from joint fits to the relevant principal correlators.
We determine the anisotropies in the $D$ and $D_s$ sectors, $\xi_{D}$ and $\xi_{D_s}$, by
fitting Eq.~\ref{equ:dispersion} to the ground state ($0^-$) $D$ and $D_s$ energies with $|\vec{n}|^2 \leq 4$.
This gives reasonable fits yielding $\xi_{D} = 3.454(6)$ and $\xi_{D_s} = 3.453(3)$.
The lines in Fig.~\ref{fig:dispersion relations} are of the form of Eq.~\ref{equ:dispersion} with
$m$ fixed to the rest energy of the relevant meson and $\xi$ fixed to our 
measured $0^-$ $D$ or $D_s$ anisotropy as appropriate.
We do not see much deviation of the higher energy states from the dispersion relations
using these anisotropies.
This suggests that we are correctly describing relativistic states up to at least $|\vec{n}|^{2} = 4$
with an anisotropy not differing greatly from the measured pion and $\eta_c$ anisotropies.
This is crucial for future calculations of scattering involving open-charm mesons.

\section{Results}
\label{sec:results}

In this section we present the results of our study.  We first compare the energy levels extracted in each lattice irrep across the two volumes, then give final spin-identified spectra and comment on other lattice investigations.  We discuss the implications of our results and compare to experiment in Section~\ref{sec:interp of results}.

\subsection{Results by lattice irrep and volume comparison}
\label{subsec:results by lattice irrep and volume comparison}

\begin{center}
\begin{figure}[tb]
\includegraphics[width=\textwidth]{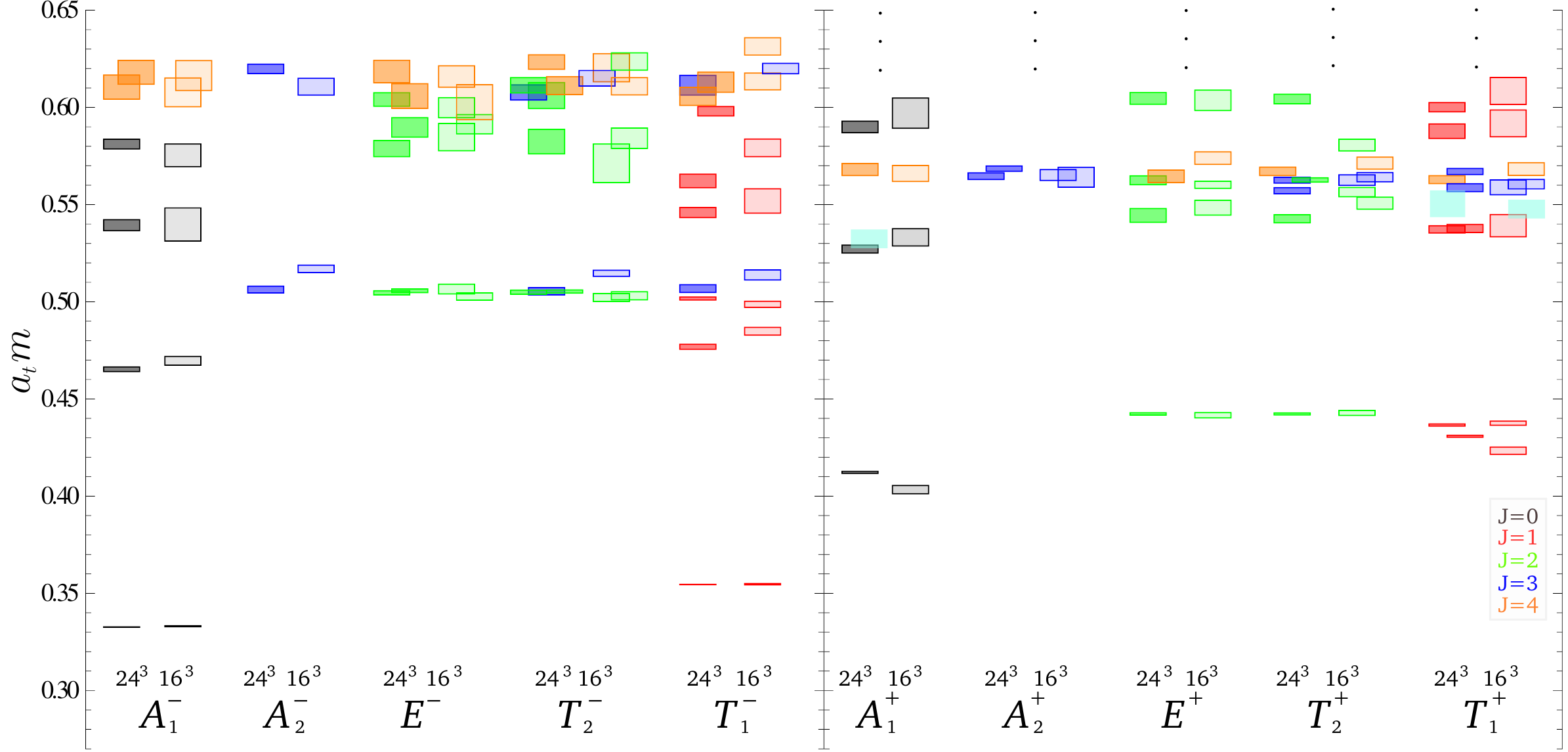}
\caption{The extracted spectrum of states in the $D$ meson sector labelled by irrep $\Lambda^{P}$. For each irrep results from both the $24^3$ and $16^3$ volumes are shown side by side. The vertical size of each box gives the one sigma statistical uncertainty on either side of the mean and the box colour refers to the continuum spin assignment as described in the text.  The light cyan boxes represent states that were not very well determined in the variational analysis; ellipses indicate that additional states may be present in these energy regions but were not robustly determined.}
\label{fig:D volume comparison by irrep}
\vspace{1cm}
\end{figure}

\end{center}

\begin{center}
\begin{figure}[tb]
\includegraphics[width=\textwidth]{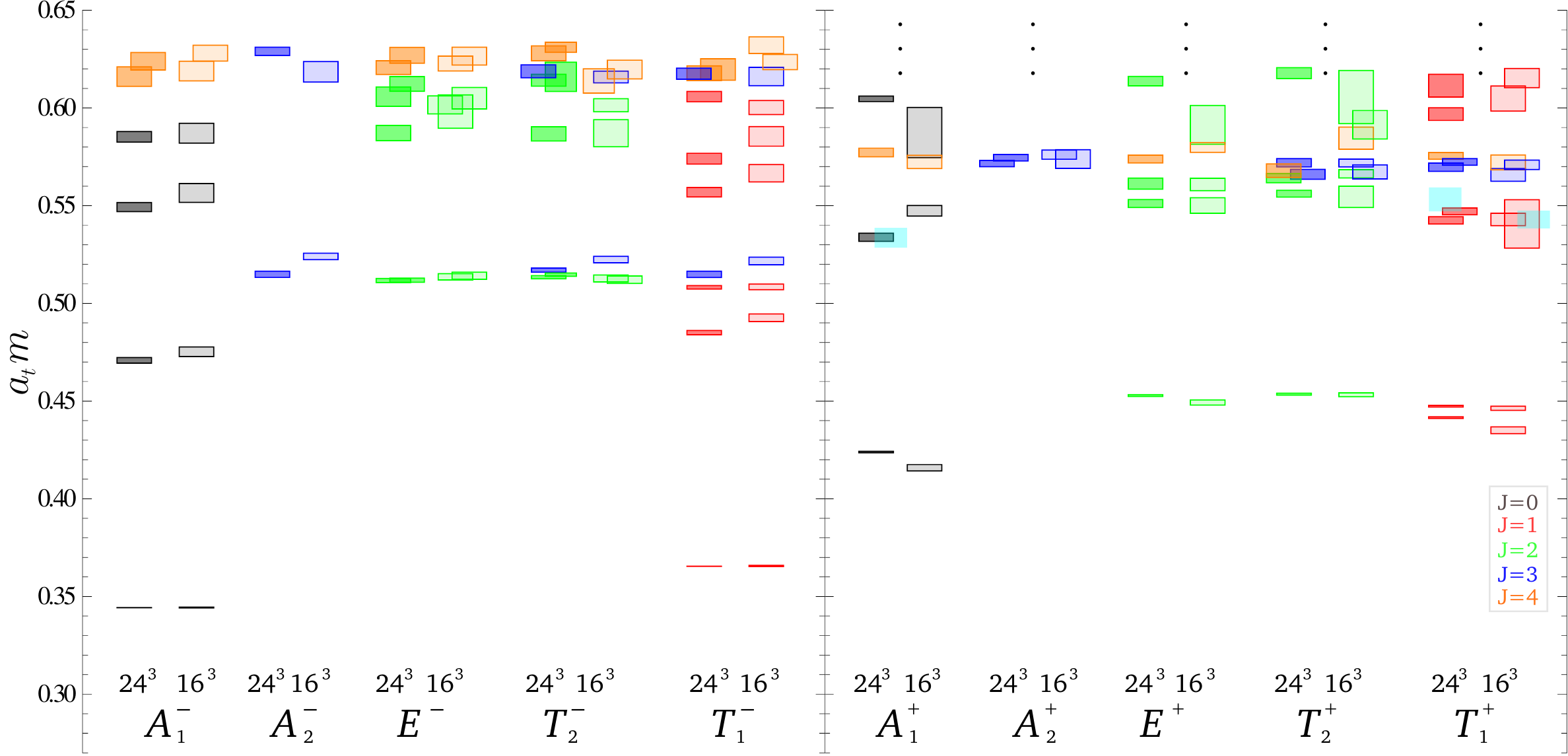}
\caption{As Fig.~\ref{fig:D volume comparison by irrep} but for the $D_{s}$ meson sector.}
\label{fig:Ds volume comparison by irrep}
\vspace{1cm}
\end{figure}
\end{center}

The results of the variational analysis for each lattice irrep in the charm-light ($D$) and charm-strange ($D_s$) sectors are shown in Figs.~\ref{fig:D volume comparison by irrep} and \ref{fig:Ds volume comparison by irrep} respectively. The vertical size of the boxes indicates the one sigma statistical uncertainty on either side of the mean while ellipses indicate that additional states may be present in the variational analysis in this energy region but these were not robust.  The colour of a box represents the continuum spin assignment from the spin identification scheme discussed in Section~\ref{subsec:spin identification}: states identified as $J = 0$ are coloured black, $J = 1$ are red, $J = 2$ are green, $J = 3$ are blue and $J = 4$ are orange.

\vspace{-2.2cm}

Components of $J \geq 2$ states are distributed across more than one lattice irrep and levels appear at similar masses, up to discretisation effects, in different irreps. The various components of the same continuum state are matched by comparing the overlaps as described in Section~\ref{subsec:spin identification}. From Figs.~\ref{fig:D volume comparison by irrep} and \ref{fig:Ds volume comparison by irrep} it is clear that there are no significant differences between energies of the same continuum state subduced into different irreps. The extremely dense spectra of states appearing at and above $a_{t}m \sim 0.52$ in the figures would be impossible to disentangle using only the extracted masses, emphasising the value of our spin identification scheme.

In general, throughout our spectra we find no significant volume dependence. However, there are a couple of exceptions: for the lightest $0^{+}$ and $1^{+}$, determined with high statistical precision, we find a $2\sigma$ discrepancy between the two volumes.  These states lie precariously close to thresholds and therefore mixing with multi-meson states may be important and could be the cause of the observed volume dependence.  Our operator bases do not include any operators that `look like' two-mesons and so, as discussed in Refs.~\cite{Dudek:2010,Dudek:2012xn}, we do not expect to be able to reliably extract multi-meson energy levels; a conservative approach is to suggest that our mass values are accurate only up to the hadronic width.  Note that in the charm-strange sector these states correspond to the enigmatic $D^{*}_{s0}(2317)^{\pm}$ and $D_{s1}(2460)^{\pm}$ which have been suggested to be molecular states of two mesons or have their properties modified by interaction with the nearby thresholds. To investigate this in more detail, in future work we will include multi-meson operators in our bases. This will result in a denser spectrum that can be analysed along the lines described in Refs.~\cite{Dudek:2012a,Dudek:2012xn}. Ref.~\cite{Mohler:2012na} has recently performed some investigations of the lightest $0^+$ and $1^+$ charm-light resonances in simulations with dynamical up and down quarks (strange-quark degrees of freedom were not included).

With these possible exceptions we find no significant volume dependence and the pattern of states across irreps is consistent with components subduced from states of definite continuum spin (multi-hadron states would show a different pattern).  We therefore generally see no clear evidence for multi-hadron states in our spectra.

\subsection{Final spin-identified spectra}
\label{subsec:final spectra}

\begin{center}
\begin{figure}[tb]
\includegraphics[width=1.0\textwidth]{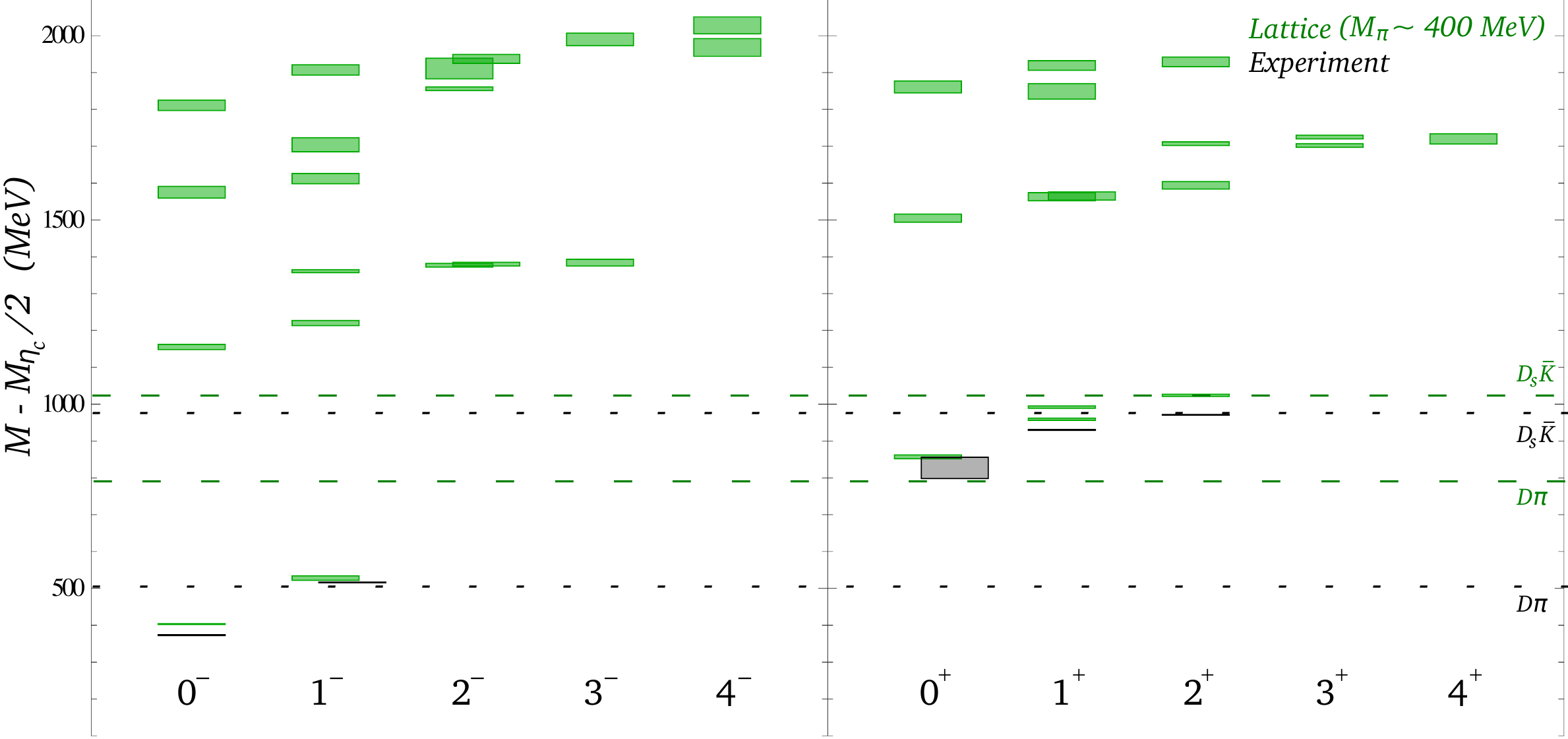}
\caption{The $D$ meson spectrum up to around $3.8$ GeV labelled by $J^{P}$. The green boxes are our calculated masses on these ensembles with $M_{\pi} \sim 400$ MeV, while the black boxes correspond to experimental masses of neutral $D$ mesons from the PDG summary tables~\cite{PDG:2012}. We present the calculated (experimental) masses with half the calculated (experimental) $\eta_{c}$ mass subtracted to reduce the uncertainty from tuning the bare charm-quark mass.  The vertical size of each box indicates the one sigma statistical uncertainty on either side of the mean. The dashed lines show the lowest non-interacting $D\pi$ and $D_{s}\bar{K}$ thresholds using our measured masses (coarse green dashing) and experimental masses (fine black dashing).}
\label{fig:D_physical}
\end{figure}
\end{center}

\begin{center}
\begin{figure}[tb]
\includegraphics[width=1.0\textwidth]{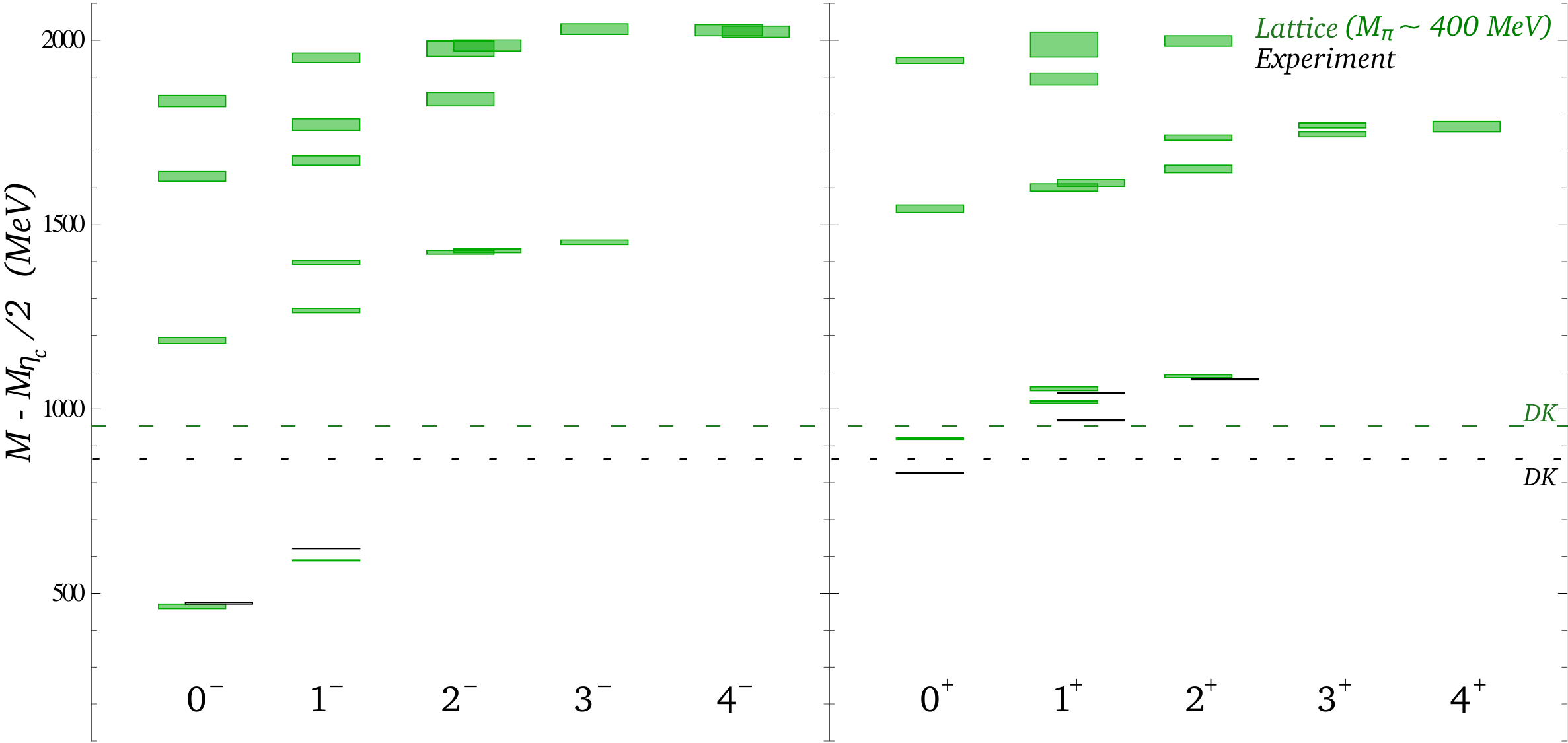}
\caption{As Fig.~\ref{fig:D_physical} but for the $D_{s}$ meson sector. The dashed lines indicate the lowest non-interacting $DK$ threshold using our measured masses (coarse green dashing) and using experimental masses (fine black dashing).}
\label{fig:Ds_physical}
\vspace{1cm}
\end{figure}
\end{center}

Our final results are the spin-identified spectra from the $24^3$ volume --
these were obtained with higher statistics than those from the $16^3$ volume and
we can reliably extract and identify the spin of a larger number of states. In
addition, the large volume means that any possible finite-volume effects will be
less important. We show our spin-identified spectra for charm-light ($D$) and
charm-strange ($D_{s}$) mesons in Figs.~\ref{fig:D_physical} and
\ref{fig:Ds_physical} respectively; here we only show well determined states for
which the spin could be reliably determined. We show the calculated
(experimental) masses with half the calculated (experimental) $\eta_{c}$ mass
subtracted in order to reduce the systematic error from any imprecision in the 
tuning of the bare charm-quark mass (see Section \ref{sec:lattice calculation}).
In Fig.~\ref{fig:D_physical}, the dashed lines correspond to the lowest non-interacting $D\pi$ and $D_{s}\bar{K}$ thresholds using our calculated masses (coarse green dashing) and using experimental masses (fine black dashing). 
The dashed lines in Fig.~\ref{fig:Ds_physical} correspond to the lowest non-interacting $DK$ threshold using our calculated masses (coarse green dashing) and using experimental masses (fine black dashing). These results are tabulated in Appendix~\ref{app:resultstable}.

\vspace{-2.0cm}

In the next section we give some interpretation of our results and compare with the experimental situation but first we discuss other lattice computations of excited charm-light and charm-strange spectra.  Most lattice studies have concentrated on the lowest-lying $S$ and $P$-wave states below threshold (see Refs.~\cite{Namekawa:2011wt,McNeile:2012qf,Dowdall:2012} for some recent studies) and only in the last couple of years have calculations begun to explore states higher up in the spectrum~\cite{Mohler:2011,Mohler:2012na,Bali:2011,Bali:2012}.  These computations, like ours, have various systematic uncertainties which have not been fully accounted for and so a direct quantitative comparison is not appropriate.  Depending on the study, these systematics come from having a finite lattice spacing, working in a finite volume, having light quarks with unphysical masses and not taking into account the finite width of states.

In Ref.~\cite{Mohler:2011} the focus is on the lowest-lying $S$ and $P$-wave multiplets though the first excited $S$-wave multiplet of $D_s$ mesons is also presented.  Dynamical ($N_{f}=2+1$) ensembles were used on a single volume for a number of different pion masses from 702~MeV down to 156~MeV; the Fermilab method~\cite{ElKhadra:1997} was used for the charm quark.  Ref.~\cite{Mohler:2012na} presents results from an $N_f=2$ calculation (i.e. with no strange-quark fields) of the charm-light spectrum on a single volume with $M_\pi = 266$ MeV using the Fermilab method for the charm quark with the correlation functions computed using the distillation technique. The lowest-lying $0^{\pm}$, $1^{\pm}$, $2^{\pm}$ states are obtained along with some excited $0^-$ and $1^-$ states, and the lightest resonances in the $0^+$ and $1^+$ channels are investigated.

Refs.~\cite{Bali:2011,Bali:2012} present preliminary results from dynamical computations of charm meson spectra for $J \leq 3$ at the $SU(3)$ point (strange quarks degenerate with the up and down quarks) with $M_\pi = 442$~MeV using the SLiNC action on a single volume. In all $J^P$ channels they obtain a ground state and a first excited state while for $0^-$ and $0^+$ they also calculate a second excited state. The treatment of the strange quarks in these calculations is very different from our approach.  The preliminary computations have $SU(3)$ flavour symmetry with $M_K = M_\pi = 442$~MeV and therefore it is not straightforward to directly compare with our results.

The operator bases used in these studies are significantly smaller than ours, they do not include as wide a range of spatial structures and in general not all the relevant lattice irreps are considered.  This means that a more limited number of states can be reliably extracted and makes a robust identification of the spin of excited and high-spin states difficult.  In addition, as discussed in the following section, our inclusion of operators proportional to the commutator of gauge-covariant derivatives allows us to explore the hybrid content of our spectra.

\section{Interpretation of results and hybrid phenomenology}
\label{sec:interp of results}

In this section we give some interpretation of our spectra. We then discuss and compare our results to the experimental situation and explore the mixing between spin-singlet and spin-triplet axial-vector, tensor and vector hybrid states.

Our final results are shown in Figs.~\ref{fig:D_physical} and \ref{fig:Ds_physical} for the charm-light and charm-strange mesons respectively, with both spectra showing a similar pattern of states. Most of the states fit into the $n^{2S+1}L_{J}$ classification expected by quark potential models, where $S$ is the spin of the quark-antiquark pair, $L$ is the relative orbital angular momentum, $n$ is the radial quantum number and $J$ is the total spin of the meson.  Following Refs.~\cite{Dudek:2008sz,Dudek:2011b}, the operator-state overlaps, $Z$, are used to identify the structure of extracted states and make $^{2S+1}L_{J}$ assignments.

In both spectra, in the negative-parity sector we find a ground state $S$-wave pair $[0^{-}, 1^{-}]$ along with a first excitation $\sim 700$ MeV higher. We see a full $D$-wave $[(1, 2, 3)^{-}, 2^{-}]$ set at $\Delta_E \sim 1400$ MeV, where $\Delta_E$ is the mass with half the $\eta_c$ mass subtracted. We observe the second excitation of the $S$-wave at $\Delta_E \sim 1900$ MeV and what appear to be parts of an excited $D$-wave set and parts of a $G$-wave $[(3, 4, 5)^{-}, 4^{-}]$ at $\Delta_E \sim 2000$ MeV. The $5^{-}$ state needed to complete the $G$-wave multiplet is not obtained in our calculation; to reliably extract and identify a spin-five state we would need to include additional operators that overlap with such a state in the continuum, requiring operators with at least four derivatives. We discuss below the supernumerary states appearing at energies between the $D$-wave and second excitation of the $S$-wave.

Moving to the positive-parity sector, we observe a full $P$-wave set $[(0,1,2)^{+}, 1^{+}]$ around the $D_{s}\bar{K}$ threshold in the charm-light spectrum and around the $DK$ threshold in the charm-strange spectrum. We find a first excited $P$-wave set $\sim 600$ MeV higher in energy. A full $F$-wave set $[(2, 3, 4)^{+}, 3^{+}]$ is present at $\Delta_E \sim 1700$ MeV in both spectra.  We discuss below the positive-parity states appearing around $\Delta_E = 1900$ MeV.

\begin{center}
\begin{figure}[tb]
\includegraphics[width=1.0\textwidth]{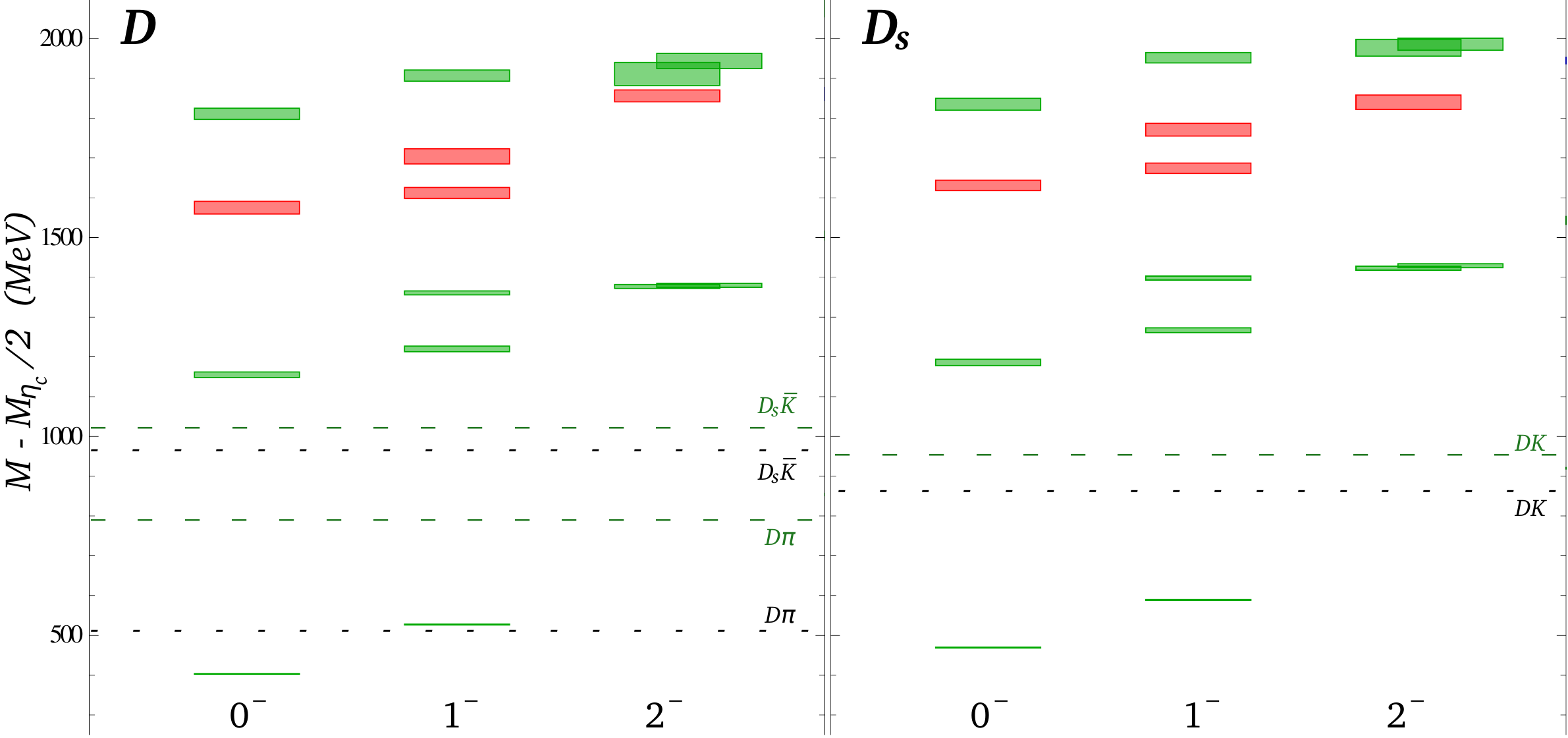}
\caption{The negative-parity $D$ (left panel) and $D_s$ (right panel) meson spectra showing only channels where we identify hybrid candidates. The red boxes are identified as states belonging to the lightest hybrid supermultiplet as discussed in the text and other notation is as in Figs.~\ref{fig:D_physical} and \ref{fig:Ds_physical}.}
\label{fig:hybrids}
\end{figure}
\end{center}
\vspace{-1.3cm}

\begin{figure}[tb]
\centering
\includegraphics[width=0.75\textwidth]{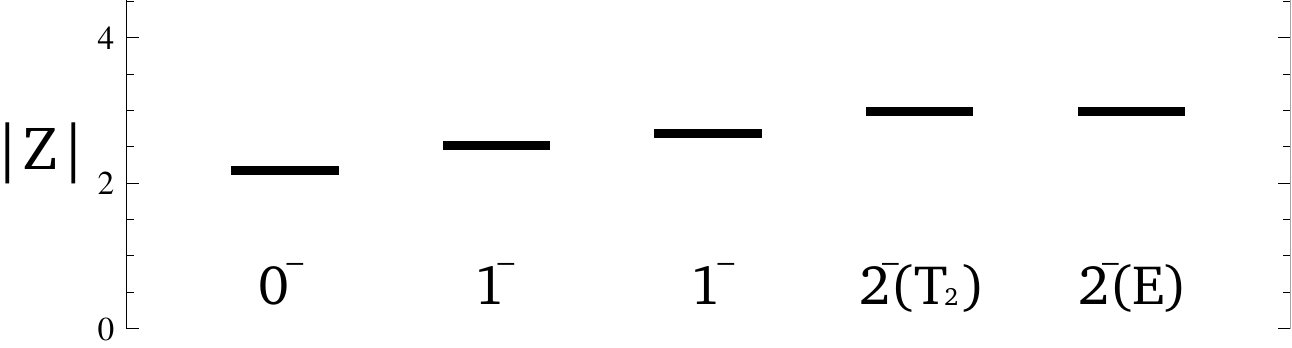}
\caption{Overlaps, $Z$, of the $D_s$ states proposed to be members of the lightest hybrid supermultiplet with operators that are proportional to the field strength tensor, $\left[\{\pi, \rho\}\times D^{[2]}_{J=1}\right]_J$ subduced into the relevant irreps.}
\label{fig:hybrid_Z}
\end{figure}

In both the charm-light and the charm-strange spectra, in the negative parity
sector we observe four states $\sim 1200$ MeV above the lowest $S$-wave states
that do not appear to fit into the pattern expected by quark models; they are
highlighted in red in Fig.~\ref{fig:hybrids}. Unlike other states, these have
relatively strong overlap with operators proportional to the field strength
tensor on smeared gauge fields, suggesting that the hadronic-scale gluonic 
field plays an important role, and so following Ref.~\cite{Dudek:2010,Dudek:2011b} we identify these as the lightest ``supermultiplet'' of hybrid mesons. As explained in Ref.~\cite{Dudek:2011b}, states within a given supermultiplet should have similar operator-state overlaps. In Fig.~\ref{fig:hybrid_Z} we show the overlaps for these states with operators $\left[\{\pi, \rho\}\times D^{[2]}_{J=1}\right]_J$. As discussed in Section~\ref{subsec:mixing}, the two $1^-$ hybrids are mixtures of spin-singlet and spin-triplet basis states and so they overlap with both the $\pi \times D^{[2]}_{J=1}$ and $\rho \times D^{[2]}_{J=1}$ operators. Therefore, for these states we plot $\sqrt{Z_\pi^2 + Z_\rho^2}$ where $Z_\pi$ and $Z_\rho$ are the overlaps with these two different operators.
The similar $Z$ values suggest that these four states have a common structure, supporting their identification as members of a supermultiplet.

The observed pattern of states in the supermultiplet, expected if a quark-antiquark pair in $S$-wave is coupled to a $1^{+-}$ excited gluonic field, and its appearance at an energy scale $\sim 1.2$ GeV above the lightest conventional state are consistent with what was found in the light meson~\cite{Dudek:2011b}, light baryon~\cite{Dudek:2012} and charmonium~\cite{Liu:2012} sectors, suggesting common physics. Those studies also discuss the first excited hybrid supermultiplet. We find candidate positive-parity hybrid states in both the charm-light and charm-strange spectra $\sim 1.5$ GeV above the ground state, the highest four positive-parity states in Figs.~\ref{fig:D_physical} and \ref{fig:Ds_physical}, but we do not robustly determine all the states in that energy region.

We have given an interpretation of our results in terms of non-relativistic
quark model multiplets.  In the heavy-quark limit, where the charm quark is much
heavier than the light and strange quarks, another classification scheme is
convenient because the spin of the heavy quark decouples.  The light degrees of
freedom have total angular momentum $j = L \otimes s_q$, where $s_q=1/2$ is the
spin of the light quark, and for $L>0$ this results in two possible values, $j =
|L \pm 1/2|$.  Coupling with the spin of the heavy quark gives a pair of values
for $J$ for each value of $j$; this pair corresponds to a degenerate doublet of
states in the heavy-quark limit, labelled by $j^P$.  For example, for $S$-wave ($L=0$) states there is one doublet with $j^P = \frac{1}{2}^-$ [$(0,1)^-$] and for $P$-wave ($L=1$) states there are two doublets, $\frac{1}{2}^+$ [$(0,1)^{+}$] and $\frac{3}{2}^+$ [$(1,2)^{+}$].  We discuss the heavy-quark limit in the context of mixing between spin-singlet and spin-triplet states in Section.~\ref{subsec:mixing}.  Although the charm quark is heavy and so these considerations may provide a useful guide, we do not expect it to be heavy enough for the heavy-quark limit to apply rigorously.  This is observed in the spectrum where, for example, the $0^-$ and $1^-$ are not degenerate.

Applying an analogous argument to hybrids, coupling the light quark degrees of freedom with a $1^+$ excited gluonic field we have $j = L \otimes s_q \otimes 1$.  For the lightest supermultiplet (with $L=0$) this gives a $j^P = \frac{1}{2}^-$ doublet with $J^P = (0,1)^-$ and a $j^P =\frac{3}{2}^-$ doublet with $J^P = (1,2)^-$.  In our spectra we note that, apart from the charm-light $2^-$ state, the $(0,1)^-$ and $(1,2)^-$ pairs of hybrids are more closely degenerate within a pair than between pairs, as suggested by the preceding argument.

\subsection{Comparison with experiment}
\label{subsec:comparison}

We now compare our results to the current experimental situation. Masses from the summary tables of the PDG review~\cite{PDG:2012} are shown in Figs.~\ref{fig:D_physical} and \ref{fig:Ds_physical} where the pattern of the rather limited number of experimental states is seen to be in general qualitative agreement with our spectra. However, some comments on the quantitative differences are warranted.

As discussed in Section~\ref{sec:results}, states above threshold can have large
hadronic widths and a conservative approach is to only consider our masses
accurate up to this scale.  An additional systematic uncertainty arises from
having light quarks that are unphysically heavy leading to a pion mass of $\sim
400$ MeV (our strange quarks have approximately the correct mass).  Below
threshold we would expect this to be relatively unimportant in the charm-strange sector, but near or above thresholds involving mesons containing light quarks this could be significant.  Another source of systematic uncertainty arises from having a finite lattice spacing and we discuss this in more detail below.  We argued in Section~\ref{sec:results} that, with a few exceptions, we see no significant finite-volume effects in our results.

We obtain a value of $124(1)$ MeV for the charm-light $S$-wave hyperfine splitting which is 18 MeV lower than the experimental value of $142.12 \pm 0.07$~MeV for the neutral mesons\cite{PDG:2012}. We get $120(1)$~MeV for the analogous charm-strange splitting which deviates by 24 MeV from the experimental value of $143.8 \pm 0.4$~MeV. In our recent study of the charmonium spectrum~\cite{Liu:2012} we found a short-fall of $\sim 40$ MeV in the $S$-wave hyperfine splitting. Investigating the dependence of the splittings on the spatial clover coefficient, $c_{s}$, we found that an increase of this coefficient brought the hyperfine splittings in line with experiment. This suggested that the discrepancy was due to lattice discretisation effects and gave an approximate scale of $\sim 40$ MeV for the leading $O(a_s)$ systematic uncertainty. We performed an analogous test for the ground state $S$-wave charm-light and charm-strange mesons and found a similar result; boosting the clover coefficient to $c_{s}=2$ from our tree-level tadpole-improved value ($c_{s}=1.35$) gave $S$-wave hyperfine splittings of $144(1)$ MeV in the charm-light sector and $139(1)$ MeV in the charm-strange sector, in line with the experimental values. This suggests that we can follow the arguments of Ref.~\cite{Liu:2012} and assign a scale of $\sim 20$ MeV to our leading $O(a_s)$ systematic uncertainty. 

In the charm-light case, we find our $P$-wave states are heavier than the
experimental equivalents; this could be due to the unphysically heavy mass of
our light quarks and/or interaction with the nearby thresholds. In the charm-strange sector, two of our $P$-wave states are consistent with experiment but the other two states, expected to correspond to the enigmatic $D^{*}_{s0}(2317)^{\pm}$ and $D_{s1}(2460)^{\pm}$, are significantly higher than their experimental counterparts. We note that the $0^+$ and $1^+$ are very close to, respectively, the $DK$ and $D^* K$ thresholds, and both the experimental and calculated $0^+$ states lie the same distance from their appropriate thresholds.  This may suggest that the unphysically heavy light quarks are a major contribution to the discrepancy. However, because of the interaction with the threshold, as discussed in Section \ref{sec:results}, further study is required with multi-hadron operators included in our bases.

We note that there have been experimental observations of additional higher-lying states; these are presented in the PDG's meson listings but omitted from the summary tables~\cite{PDG:2012}. Because they are still awaiting experimental confirmation we do not include them in our plots.

\subsection{Mixing of spin-singlet and triplet states}
\label{subsec:mixing}

In QCD, the charm quark is significantly heavier than the strange and light (up, down) 
quarks and therefore $SU(4)$ flavour symmetry is badly broken.  In
particular, charm-strange and charm-light mesons are not eigenstates of a
generalisation of charge-conjugation.  In contrast, $SU(2)$ isospin symmetry is
a reasonable approximation in the light quark sector and light mesons are
eigenstates of $G$-parity to a good approximation\footnote{Our lattice
  calculations have degenerate up and down Wilson quarks so isospin is an exact symmetry.}.  The absence of such a symmetry for charm mesons means that $S=1$ and $S=0$ quark-model states with $J=L$ ($^3L_{J=L}$ and $^1L_{J=L}$) can mix and probing this mixing can help us understand the internal dynamics and quantify flavour symmetry breaking.  For flavourless mesons these would correspond to $J^{PC} = J^{\pm\pm}$ and $J^{\pm\mp}$ respectively. Using a two-state hypothesis and assuming energy-independent mixing (the states are close in energy), we can expand the pair of states $A$ and $B$ (we have chosen $B$ to be the heavier state of our pair) in terms of spin-singlet and triplet basis states,
\begin{eqnarray}
 \rvert A \rangle &=& + \cos \theta ~ \rvert ^{1}L_{J=L} \rangle + \sin \theta ~ \rvert ^{3}L_{J=L} \rangle ~, \nonumber \\
 \rvert B \rangle &=& - \sin \theta ~ \rvert ^{1}L_{J=L} \rangle + \cos \theta ~ \rvert ^{3}L_{J=L} \rangle ~.
\end{eqnarray}
In the non-relativistic limit, our operator $\left[(\rho - \rho_{2}) \times D^{[L]}_L \right]_{J=L}$ only overlaps onto $^3L_{J=L}$ states and $\left[\{\pi, \pi_{2}\} \times D^{[L]}_L\right]_{J=L}$ only overlap onto $^1L_{J=L}$ states.  There are analogous operators, $\left[(\rho - \rho_{2}) \times D^{[2]}_1 \right]_{J=1}$ and $\left[\{\pi, \pi_{2}\} \times D^{[2]}_1\right]_{J=1}$, which overlap onto respectively the spin-triplet and spin-singlet $1^-$ hybrids.  To determine the mixing angle $\theta$, we take the ratio of the overlap factor of one of the aforementioned operators for state $A$ to the overlap factor for state $B$; this ratio gives $\tan \theta$ or $\cot \theta$ depending on the operator used.

\begin{table*}[tb]
\begin{center}
\begin{tabular}{|c|c|c c c|c|}
\hline
 & & \multicolumn{4}{|c|}{$|\theta| /^{\circ}$} \\
 & $J^{P}$ & $\sim \left(\rho - \rho_{2}\right)$ & $\sim \pi$  & $\sim \pi_{2}$ & Heavy-quark limit \\
\hline
 c-l & $1^{+}$           & 60.1(0.4) & 62.6(0.2) & 65.4(0.2) & 54.7 or 35.3 \\
     & $2^{-}$           & 26.7(2.2) & 22.2(3.7) & 18.9(3.9) & 50.8 or 39.2 \\
     & $1^{-}$ (hybrid)  & 59.7(1.1) & 68.4(0.8) & 67.4(0.9) &  \\
\hline
 c-s & $1^{+}$           & 60.9(0.6) & 64.9(0.2) & 66.4(0.4) & 54.7 or 35.3 \\
     & $2^{-}$           & 64.9(1.9) & 68.7(2.0) & 70.9(1.8) & 50.8 or 39.2 \\
     & $1^{-}$ (hybrid)  & 59.9(1.7) & 67.9(0.9) & 67.3(0.9) & \\
\hline
\end{tabular}
\caption{The absolute value of the mixing angles for the lightest pairs of $1^+$, $2^-$ and hybrid $1^-$ states in the charm-light (c-l) and charm-strange (c-s) sectors. The angles extracted using different operators are presented; these are labelled by the gamma matrix structure with the derivative structures described in the text.  Also shown are the mixing angles expected in the heavy-quark limit. The apparent difference between the charm-light and charm-strange $2^-$ mixing angles is explained in the text.}
\label{table:mixing angles}
\end{center}
\end{table*}

Extracted mixing angles for the lightest pairs of $P$-wave ($1^+$), $D$-wave ($2^-$) and hybrid ($1^-$) states are presented in Table~\ref{table:mixing angles} for each of the operators considered. Note that the overall sign (or in general the overall phase) of the overlaps for each state is not observable in our calculation and hence we determine the absolute value of the mixing angles. 
For each pair of states, the variation between mixing angles determined using the three different operators gives an idea of the size of the systematic uncertainties. As well as the usual lattice systematics, these include uncertainties arising from this simple interpretation, a two-state hypothesis with energy-independent mixing, and from assuming that the energy difference between the states is sufficiently small such that the renormalisation factors for the operators do not vary substantially over this energy range.
A relatively large variation is seen in the $1^-$ hybrid mixing angles; this would be expected because here the
assumptions are less justified: these states lie higher in the spectrum, there are
other states relatively close by with the same quantum numbers and there is a relatively large mass splitting within a pair.

The $P$-wave ($^3P_{1}$ - $^1P_{1}$) mixing angles are similar in the charm-strange and charm-light sectors, as are the mixing angles for the vector hybrids. 
For the $D$-waves ($^3D_{2}$ - $^1D_{2}$), $\theta_{c-s} \approx 90^{\circ} - \theta_{c-l}$ because the dominantly $^3D_{2}$ state and the dominantly $^1D_{2}$ state are almost degenerate, and the mass ordering of our states is reversed in the charm-strange sector compared to the charm-light sector.
Although our strange quark mass is close to the physical value~\cite{Edwards:2008}, our light quarks are unphysically heavy giving a pion mass of $\sim 400$ MeV. Therefore $SU(3)$ (strange-light) flavour symmetry is not badly broken (e.g. $M_K/M_\pi$ is close to unity~\cite{Dudek:2010}) and we would expect to find similar mixing angles in the charm-light and charm-strange sectors. Because we have unphysically heavy light quarks, our determinations of mixing angles in the charm-strange sector might be expected to be closer to the physical values than those in the charm-light sector, but we note the other sources of systematic uncertainty discussed above.

In Table~\ref{table:mixing angles} we also show the mixing angles expected in
the heavy-quark limit where $m_c \gg m_{u,d,s}$~\cite{Close:2005}.  As discussed above, in this limit the heavy-quark spin decouples and states appear in doublets labelled by $j^P$ where $j$ is the total angular momentum of the light degrees of freedom.  For $P$-wave states there are two doublets $\frac{1}{2}^+$ [$(0,1)^{+}$] and $\frac{3}{2}^+$ [$(1,2)^{+}$]. The $j^P$ $1^+$ states (appropriate for the heavy-quark limit) are linear combinations of the singlet ($^{1}P_{1}$) and triplet ($^{3}P_{1}$) states (appropriate for the equal-mass limit) corresponding to a particular mixing angle. An analogous situation arises for the other pairs of states.

Our mixing angles lie between zero mixing (the flavour symmetry limit, i.e. $0^{\circ}$ or $90^{\circ}$) and the heavy-quark limit values. This is consistent with what one expects because the charm quark mass is at an intermediate scale, larger than the light and strange quark masses but not large enough for the heavy-quark limit to apply. Quark model calculations and other determinations~\cite{Godfrey:1985,Godfrey:1991,Abe:2003zm,Badalian:2007yr,Badalian:2011tb,Wu:2011yb} in general find significant mixing between the spin-singlet and spin-triplet states with, at least for the lightest $1^+$ pair, some deviation from the heavy quark limit.  In comparison, because $SU(3)$ flavour symmetry is not strongly broken in our calculations, near zero mixing was found in the kaon sector~\cite{Dudek:2010}; we would expect the analogous mixing angles in the $B$ meson system to be closer to the heavy-quark limit values.

\section{Summary}
\label{sec:Summary}

We have computed extensive spectra of charm-light and charm-strange mesons using distillation and the variational method combined with a large basis of carefully constructed operators. Along with our spin
identification scheme, these techniques have allowed us to extract a high number of states across all possible $J^{P}$ combinations, up to and including states of spin four. Our calculations were performed on two volumes and, apart from the exceptions mentioned earlier, we do not observe any significant volume dependence.

Much of our spectra appear to follow the $n^{2S+1}L_{J}$ pattern expected by quark models but we also observe an excess of states that do not appear to fit into such a scheme. We have suggested that these
states be interpreted as hybrid mesons due to their strong gluonic contribution. Their pattern ($J^{P} = [(0, 1, 2)^{-}, 1^{-}]$) and energy scale are consistent with that previously observed in both the light meson~\cite{Dudek:2011b} and charmonia~\cite{Liu:2012} sectors, and can be interpreted as a colour-octet quark-antiquark pair coupled to a $1^{+-}$ chromomagnetic excitation.

Using a non-relativistic interpretation of some of our operators, we extracted mixing angles between the spin-singlet and spin-triplet axial-vector $1^{+}$ states, $2^{-}$ $D$-wave states and $1^{-}$ hybrid states. We found mixing angles intermediate between zero mixing and the heavy quark limit values. Results in both the charm-light and charm-strange sectors are very similar, but we note that $SU(3)$ flavour symmetry breaking is suppressed on our ensembles as we have unphysically-heavy up and down quarks; $M_{K}$/$M_{\pi}$ is close to unity. Nevertheless, we expect the flavour symmetry breaking in the charm-strange sector to be of the correct scale, bearing in mind the other systematic uncertainties discussed above.

A full understanding of the charm sector requires us to study resonances and this is particularly relevant for many of the enigmatic charmed and charmonium-like states. We have calculated dispersion relations for the low-lying charm-light and charm-strange states, laying the foundations for a study of scattering states, along the lines of Refs.~\cite{Dudek:2012a,Dudek:2012xn}, using carefully constructed multi-hadron operators and the L\"{u}scher methodology~\cite{Luscher:1990} with its extensions.

\begin{acknowledgments}
We thank our colleagues within the Hadron Spectrum Collaboration.  
{\tt Chroma}~\cite{Edwards:2004sx} and {\tt QUDA}~\cite{Clark:2009wm,Babich:2010mu} were used to perform 
this work on the Lonsdale cluster maintained by the Trinity Centre for High Performance Computing 
funded through grants from Science Foundation Ireland (SFI), 
at the SFI/HEA Irish Centre for High-End Computing (ICHEC), 
and at Jefferson Laboratory under the USQCD Initiative and the LQCD ARRA project.  
Gauge configurations were generated using resources awarded from the U.S. Department of Energy 
INCITE program at the Oak Ridge Leadership Computing Facility at Oak Ridge National Laboratory, the NSF Teragrid at the Texas Advanced Computer Center 
and the Pittsburgh Supercomputer Center, as well as at Jefferson Lab.  
This research was supported by the European Union under Grant Agreement 
number 238353 (ITN STRONGnet) and by the Science Foundation Ireland under Grant Nos.~RFP-PHY-3201 and RFP-PHY-3218.
GM acknowledges support from the School of Mathematics at Trinity College Dublin.
CET acknowledges support from a Marie Curie International Incoming Fellowship, 
PIIF-GA-2010-273320, within the 7th European Community Framework Programme. 
\end{acknowledgments}

\appendix
\section{Tables of results}
\label{app:resultstable}

In Tables~\ref{table:D results} and \ref{table:Ds results} we tabulate,
   respectively, the charm-light ($D$) and charm-strange ($D_s$) masses
   calculated on the $24^3$ ensemble, as presented in Figs.~\ref{fig:D_physical}
   and \ref{fig:Ds_physical}.  The scale has been set using the $\Omega$-baryon
   mass as described in Section~\ref{sec:lattice calculation}; note that $M_\pi
   \sim 400$~MeV in these computations.  Half the calculated mass of the
   $\eta_c$ has been subtracted in order to reduce the systematic uncertainty
   arising from any imprecision in tuning the bare charm-quark mass (see Section \ref{sec:lattice calculation}). 

\begin{table}[htb]
\begin{center}
\begin{tabular}{|c|llllll|}
\hline
$J^{P}$ &\multicolumn{6}{|c|}{$M-M_{\eta_c}/2$  (MeV)}\\
\hline
$0^{-}$ &403(1)&1155(7)&1575(16)&1811(14)&& \\
$1^{-}$ &527(1)&1220(7)&1361(5)&1612(14)&1704(19)&1907(14) \\
$2^{-}$ &1377(5)&1380(5)&1856(15)&1911(29)&1944(19)& \\
$3^{-}$ &1384(9)&1990(17)&&&& \\
$4^{-}$ &1968(24)&2028(23)&&&& \\
\hline
$0^{+}$ &854(3)&1505(11)&1861(16)&&& \\
$1^{+}$ &959(3)&992(3)&1563(11)&1565(11)&1849(21)&1919(13) \\
$2^{+}$ &1024(3)&1594(10)&1707(5)&1929(13)&& \\
$3^{+}$ &1708(6)&1718(8)&&&& \\
$4^{+}$ &1720(14)&&&&& \\
\hline
\end{tabular}
\caption{Summary of the $D$ meson spectrum, as presented in Fig.~\ref{fig:D_physical}, with statistical uncertainties shown.  Note that $M_\pi \sim 400$~MeV in these computations and more details are given in the text.} 
\label{table:D results}
\end{center}

\end{table}
\begin{table}[htb]
\begin{center}
\begin{tabular}{|c|llllll|}
\hline
$J^{P}$ &\multicolumn{6}{|c|}{$M-M_{\eta_c}/2$  (MeV)}\\
 \hline
$0^{-}$ &469(1)&1186(8)&1631(13)&1835(15)&& \\
$1^{-}$ &589(5)&1267(6)&1398(5)&1674(13)&1771(16)&1952(13) \\
$2^{-}$ &1425(4)&1429(4)&1840(18)&1977(21)&1986(15)& \\
$3^{-}$ &1452(6)&2030(14)&&&& \\
$4^{-}$ &2023(14)&2027(16)&&&& \\
\hline
$0^{+}$ &920(2)&1543(10)&1945(8)&&& \\
$1^{+}$ &1019(3)&1055(5)&1601(10)&1613(9)&1895(16)&1988(34) \\
$2^{+}$ &1087(2)&1651(10)&1736(7)&1998(14)&& \\
$3^{+}$ &1749(7)&1767(7)&&&& \\
$4^{+}$ &1766(14)&&&&& \\
\hline
\end{tabular}
\caption{Summary of the $D_{s}$ meson spectrum, as presented in Fig.~\ref{fig:Ds_physical}, with statistical uncertainties shown.  Note that $M_\pi \sim 400$~MeV in these computations and more details are given in the text.} 
\label{table:Ds results}
\end{center}
\end{table}


\bibliography{paper}

\end{document}